\documentclass[prb,showpacs,preprintnumbers,amsfonts,amssymb,amsmath,floats,twocolumn,aps]{revtex4}

\usepackage{graphicx}
\usepackage{dcolumn}
\usepackage{bm}
\usepackage[final,dvips]{epsfig}
\usepackage{bm}
\usepackage{color}

\newcommand{\ocite}[1]{Ref.\ \onlinecite{#1}}
\newcommand{\be}{\begin{equation}}
\newcommand{\ee}{\end{equation}}
\newcommand{\ba}{\begin{eqnarray}}
\newcommand{\ea}{\end{eqnarray}}
\newcommand{\ff}[1]{{\bm #1}}
\newcommand{\tr}{\mbox{tr}}
\newcommand{\Tr}{\mbox{Tr}}
\newcommand{\refeq}[1]{Eq.\ (\ref{eq:#1})}
\newcommand{\labeq}[1]{\label{eq:#1}}
\newcommand{\reffig}[1]{Fig.\ \ref{fig:#1}}
\newcommand{\labfig}[1]{\label{fig:#1}}
\newcommand{\rem}[1]{}

\begin{document} 
  
\title{Variational cluster approach to ferromagnetism in infinite dimensions and in one-dimensional chains}
 
\author{Matthias Balzer}
\author{Michael Potthoff}

\affiliation{I. Institut f\"ur Theoretische Physik, Universit\"at Hamburg, Jungiusstra\ss{}e 9, D-20355 Hamburg, Germany}
 
\begin{abstract}
The variational cluster approach (VCA) is applied to study spontaneous ferromagnetism in the Hubbard model at zero temperature. 
We discuss several technical improvements of the numerical implementation of the VCA which become necessary for studies of a ferromagnetically ordered phase, e.g.\ more accurate techniques to evaluate the variational ground-state energy, improved local as well as global algorithms to find stationary points, and different methods to locate the magnetic phase transition.
Using the single-site VCA, i.e.\ the dynamical impurity approximation (DIA), the ferromagnetic phase diagram of the model in infinite dimensions is worked out.
The results are compared with previous dynamical mean-field studies for benchmarking purposes. 
The DIA results provide a unified picture of ferromagnetism in the infinite-dimensional model by interlinking different parameter regimes that are governed by different mechanisms for ferromagnetic order.
Using the DIA and the VCA, we then study ferromagnetism in one-dimensional Hubbard chains with nearest and next-nearest-neighbor hopping $t_2$. 
In comparison with previous results from the density-matrix renormalization group, the phase diagram is mapped out as a function of the Hubbard-$U$, the electron filling and $t_2$. 
The stability of the ferromagnetic ground state against local and short-range non-local quantum fluctuations is discussed. 
\end{abstract} 
 



\maketitle 

\section{Introduction}

Itinerant ferromagnetism of nanometer-sized transition-metal systems deposited on non-magnetic surfaces has attracted much attention recently. 
It is a fascinating physical but also technological vision to control the geometrical arrangement of the nanosystem on the atomic scale while studying its magnetic properties with atomic resolution. \cite{Wie09}
This provokes new and exciting questions. 
It is highly interesting, for example, to understand how many atoms are necessary and how these atoms should be arranged geometrically to ensure a stable ferromagnetic state. 

One important issue for magnetic nanosystems is their stability against thermal fluctuations. \cite{Bro63}
This is mainly determined by anisotropies.
Anisotropic contributions to the total energy of a magnetic system can be several orders of magnitude higher at a surface or for a small cluster or chain as compared to a three-dimensional bulk of the same material. 
Nevertheless, the anisotropy strength is usually still much smaller than the exchange coupling and can thus be safely disregarded for the question of whether or not a ferromagnetic {\em ground state} is existing. 

The stability of a ferromagnetic ground state is, therefore, a matter of quantum fluctuations.
Opposed to antiferromagnetic order, for example, the order parameter is a conserved quantity in the case of ferromagnetism. 
It is thus mainly the quantum fluctuations of the {\em para}magnetic state which are important and which the spin-polarized state is competing with. 

The ground state of itinerant systems, \cite{BDN01} such as mono-atomic chains of 3d transition metals, \cite{GDM+02} is of particular interest from a theoretical point of view.
Besides the subtle interplay between the kinetic energy of the itinerant electrons and their Coulomb interaction, geometrical constraints come into play additionally. 
Even for a bulk system, however, and even for the most elementary models of itinerant ferromagnetism, such as the Hubbard model, \cite{Hub63,Gut63,Kan63} there is no simple and comprehensive physical picture for the mechanism that drives ferromagnetic order. \cite{NB89,HN97b,PHWN98} 

The physical reason which hampers a straightforward understanding of itinerant ground-state ferromagnetism probably consists in the fact that the ordering and actually the formation of local magnetic moments is a strong-coupling phenomenon and thus in general not capable by perturbative techniques. 
This is opposed to antiferromagnetic order, for example: \cite{Geb97}
Slater or band antiferromagnetism is accessible by weak-coupling approaches, Heisenberg or local-moment antiferromagnetism emerges in effective low-energy models.

It is therefore important to recognize that the same problems already show up in the Hubbard model on infinite-dimensional lattices.
This limit, however, is rigorously accessible by dynamical mean-field theory (DMFT), \cite{GKKR96,PJF95,KV04,Vol10} and many insights concerning itinerant ferromagnetism could be obtained in this way. \cite{VBH+99} 
First of all, quantum fluctuations are recognized as essential.
This means that a static mean-field approach, like Hartree-Fock theory, cannot grasp the main physics and largely overestimates the tendency to collective ordering.
Further, ferromagnetism requires a strong local Coulomb interaction and must therefore be investigated by non-perturbative means. \cite{OPK97,ZPB02,PHMK08}
In addition, however, subtle details of the non-interacting electronic structure are likewise important, e.g.\ a strong asymmetry of the local density of states. \cite{Ulm98,WBS+98,Uhr96} 
Non-local parts of the Coulomb interaction, a non-local ferromagnetic Heisenberg exchange coupling, for example, do affect the magnetic ground-state phase diagram but are found to be of lesser importance as compared to the Hubbard-$U$ in general. 
Finally, ferromagnetic order strongly competes with antiferromagnetism and is realized away from half-filling.

While DMFT can be regarded as the optimal theoretical framework to deal with strong local quantum fluctuations and to understand their effect on itinerant ferromagnetism, it is still a mean-field approach.
This means that the feedback of non-local two-particle, e.g.\ magnetic, excitations on the one-particle spectrum and also on the thermodynamics is neglected.
It is presently unclear, as how severe this approximation must be regarded when studying low-dimensional systems, for example. 
Spin-charge separation, \cite{Voi95} to mention a prominent example of a non-local quantum effect in one-dimensional chains, cannot be described by DMFT. 
As concerns ferromagnetic order in one-dimensional itinerant systems, however, there is reason to be more optimistic that DMFT may capture the essential physics: 
Namely, magnetic correlations and thus the feedback of magnetic correlations on the ground state can be expected to become less important for fillings well below half-filling.
At and around half-filling {\em anti}ferromagnetic correlations dominate anyway.
In fact, numerically exact studies by means of the density-matrix renormalization group (DMRG) for the $t_1$-$t_2$ one-dimensional Hubbard model \cite{DN97,DN98,Dau00} yield a ground-state ferromagnetic phase diagram which shows striking similarities with the DMFT results and confirm the main qualitative results listed above. 

This situation has motivated the present study which employs the variational cluster approach (VCA) \cite{PAD03,DAH+04} to investigate the ferromagnetic ground-state phase diagram of the infinite- and the one-dimensional Hubbard model.
The VCA is a thermodynamically consistent \cite{AAPH06a} cluster mean-field approach which determines the electron self-energy by exploiting a general variational principle. \cite{Pot03a,Pot03b,Pot05}
Different approximations can be constructed by the choice of different reference systems that define the space of test self-energies for the variational principle.
In this way, single-site mean-field approximations, very close to the DMFT, as well as cluster approximations can be constructed which include the local but also non-local quantum fluctuations, respectively.

By comparison with previous DMFT and DMRG results it should be possible to answer the following interrelated questions: 
How sensitive is a ferromagnetic state on a one-dimensional chain to local quantum fluctuations?
What affects its stability more, local or short-range non-local fluctuations?
Is a single-site mean-field approach sufficient to predict stable ferromagnetic phases?
How much does it improve compared to a purely static approach?
Does an inclusion of short-range non-local fluctuations improve the predictive power?
Answers to these questions are particularly important for future studies of magnetic nanosystems in more complex geometries such as clusters, coupled chains, etc.\ and including more orbitals per sites since those systems are in most cases not accessible to an exact numerical treatment via the DMRG.

A second and likewise important goal of our study is to advance the variational cluster approach:
Its evaluation requires the repeated calculation of the self-energy of the reference system for different one-particle parameters which serve as variational parameters.
At zero temperature the numerical solution has to be performed using exact-diagonalization techniques. 
It is then clear that the quality of the approximation is limited by the exponential growth of the reference system's Hilbert space, i.e.\ by the limited number of sites that can be taken into account.
However, an increasing number of sites in the reference system at the same time means that the number of variational parameters increases. 
To study ferromagnetic phases, the number of parameters is doubled because of the additional spin-dependence of each of the parameters. 
The variationally determined ground-state energy becomes decreasingly sensitive with each additional variational degree of freedom considered.
This tightens the need for extremely accurate computations.
Here our goal is to present and discuss different technical improvements of the VCA. 

The paper is organized as follows:
In the following section \ref{VCA} we briefly review the theoretical concept and then address the different technical issues important for a reliable numerical evaluation in Sec.\ \ref{NUM}.
Results for the Hubbard model in infinite dimensions and in one dimension are presented and discussed in Sec.\ \ref{FM}, and a summary of the main conclusions is given in
Sec.\ \ref{CON}.

\section{Variational cluster technique}
\label{VCA}

We consider the single-band Hubbard model \cite{Hub63,Gut63,Kan63} in one dimension with nearest and next-nearest neighbor hopping $t_1$ and $t_2$, respectively (except for Sec.\ \ref{sec:infd}).
Using standard notations, the Hamiltonian reads
\be
H = - t_1 \sum_{< ij >,\sigma} c^\dagger_{i\sigma} c_{j\sigma} 
    - t_2 \sum_{\ll ij \gg, \sigma} c^\dagger_{i\sigma} c_{j\sigma} 
    + U \sum_{i} n_{i\uparrow} n_{i\downarrow}
\labeq{hubbard}
\ee
where $<\cdot >$ and $\ll\cdot\gg$ restrict the independent sums over lattice sites $i$ and $j$ to nearest and next-nearest neighbors, respectively.
$\sigma=\uparrow,\downarrow$ is the spin projection. 
The strength of the on-site Hubbard interaction is given by $U$.
In the following we consider finite Hubbard chains consisting of $L$ sites and assume periodic boundary conditions. 
Unless stated differently, we set $t_1=1$ to fix the energy scale.

Calculations are performed using the variational cluster approximation \cite{PAD03,DAH+04} (VCA) which is a quantum cluster mean-field approach based on the self-energy functional theory \cite{Pot03a,Pot03b,Pot05} (SFT).
Central to the SFT is the self-energy functional
\be
\Omega[\ff \Sigma] = \Tr \ln \left(\ff G^{-1}_0 - \ff \Sigma \right)^{-1} + F[\ff\Sigma] \: ,
\labeq{SFT}	
\ee
which provides an exact functional relation between the self-energy $\ff \Sigma$ (with elements $\Sigma_{ij\sigma}(\omega_n)$) and the grand potential $\Omega$ of \refeq{hubbard} at temperature $T$ and chemical potential $\mu$.
Furthermore, $\ln$ denotes the main branch of the complex logarithm, and $\Tr \equiv T \sum_n \exp(i\omega_n 0^+) \, \tr$ with $\tr$ being the trace over the spatial degrees of freedom and $0^+$ a positive infinitesimal which ensures convergence of the sum over fermionic Matsubara frequencies $\omega_n$.
$\ff G_0$ is the free single-particle Green's function of the model which can be assumed to be known. 
The functional $F[\ff\Sigma]$ is formally defined as the Legendre transform of the Luttinger-Ward functional $\Phi[\ff G]$ which in turn is defined diagrammatically \cite{LW60} or through a functional integral. \cite{Pot06b}

The self-energy functional is constructed such that it becomes stationary at the exact (physical) self-energy of the model system \refeq{hubbard}:
\be
\delta\Omega[\ff\Sigma] = 0 \: .
\labeq{stationarity}
\ee
Due to the fact that $F[\ff \Sigma]$ is not known explicitly, an approximation must be employed to make use of this variational principle.
The idea of the SFT is to restrict the variation of the self-energy in \refeq{stationarity} to a subspace of trial self-energies which is spanned by the exact self-energies of a certain reference system. 
On this subspace, the self-energy functional can be evaluated exactly, provided that the reference system has the same interaction part as the original model and provided that an exact (numerical) computation of the test self-energies is possible (see Refs.\ \onlinecite{Pot03a,Pot03b,Pot05} for details).
Generically, a reference system is a finite Hubbard cluster with the same Hubbard-$U$ but with its one-particle parameters (hopping and on-site energies) serving to parametrize the test self-energies. 
Usually, one selects a limited number of one-particle parameters as variational parameters $\ff\lambda=(\lambda_1,\dots,\lambda_s$). 
Then, the stationarity condition \refeq{stationarity} is approximated requiring the function
\be
\Omega(\ff\lambda)\equiv\Omega[\ff\Sigma(\ff\lambda)]
\labeq{omegalambda} 
\ee
to be stationary, i.e.\ $\partial \Omega(\ff\lambda) / \partial \ff \lambda = 0$.

\begin{figure}
\centerline{\includegraphics[scale=.24,clip=]{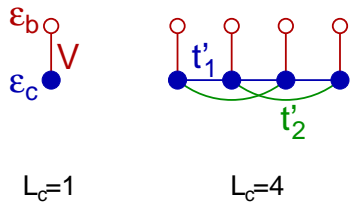}}
\caption{Examples for reference systems with $L_c$ correlated sites (blue circles) and one additional bath site (red circles) per correlated site generating single-site (mean-field) and cluster approximations.
The (spin-dependent) on-site energies $\varepsilon_c$ and $\varepsilon_b$ as well as the (spin-dependent) hybridization $V$ are optimized.
Hopping parameters $t'_1$ and $t'_2$ are kept fixed at their physical values.
}
\labfig{refsys}
\end{figure}

For the present study of the one-dimensional Hubbard model, we consider chains with $L_c \leq 5$ correlated sites as reference systems. 
One additional uncorrelated (``bath'') site is attached to each correlated site (i.e.\ $n_s=2$ local degrees of freedom), see \reffig{refsys}. 
Physically, this means to take retardation effects into account, i.e.\ with this choice of the reference system, local (temporal) fluctuations are included to some degree. 
Local fluctuations are treated {\em exactly} in the limit of $n_s=\infty$ only, i.e.\ for a continuum of bath degrees of freedom (see Ref.\ \onlinecite{Pot05}).
This would correspond to the (cellular) dynamical mean-field (C-DMFT) approach \cite{KSPB01}. 
It has been demonstrated in various contexts, \cite{Poz04,EKPV07,BHP08,BKS+09} however, that the main effect of local correlations is already accounted for with $n_s=2$, i.e.\ the essential step is the one from a plain VCA (no bath sites) to an $n_s=2$-VCA while more bath sites give secondary corrections.
This is important since only a finite (small) number of sites can be treated when using an exact-diagonalization technique at temperature $T=0$ to compute the chain self-energy.
Physically, the main point is that the $n_s=2$ reference systems already allow for a {\em local} (Kondo-type) singlet formation to screen the local magnetic moments.
A {\em non-local} singlet formation is possible for reference systems with $L_c \geq 2$. This describes the feedback of non-local magnetic correlations on the single-particle excitation spectrum.
The degree to which, quite generally, spatial correlations are accounted for by the VCA is controlled by the choice of $L_c$, ranging from a (dynamical) single-site mean-field approximation for $L_c=1$ over cluster mean-field approaches to the exact solution that is (in principle) obtained with $L_c=\infty$.

For a given reference cluster (\reffig{refsys}), we treat the on-site energies of the correlated and of the bath sites, $\varepsilon_c$ and $\varepsilon_b$, as variational parameters. 
This ensures thermodynamic consistency \cite{AAPH06a} with respect to the particle density $n$:
Even within the approximation, exactly the same result is obtained for $n$ which either can be determined via a $\mu$-derivative of the SFT grand potential at stationarity or via a frequency integral over the one-electron spectral density corresponding to $\ff G = (\ff G_0^{-1} - \ff \Sigma)^{-1}$ with the optimal self-energy. 
Furthermore, to control the temporal fluctuations, the hybridization $V$ is considered as variational parameter (see \reffig{refsys}).
For the present study it is important to allow for a possible spin dependence of all variational parameters to have thermodynamical consistency with respect to the magnetization in addition.
As a simplifying but excellent \cite{BHP08} assumption to limit the number of parameters, $\varepsilon_c$, $\varepsilon_b$ and $V$ are taken to be site independent, and the hopping parameters within the reference chain are fixed at their original values, i.e.\ $t'_1=t_1$ and $t'_2=t_2$.

Calculations are performed for the grand canonical ensemble keeping the chemical potential $\mu$ fixed.
Due to the discrete energy spectrum of the finite reference cluster, and due to the U(1) symmetry of the cluster Hamiltonian $H'$, however, the cluster ground state reveals a fixed total particle number $N'$ within finite $\mu$ ranges.
Therefore, the cluster electron density $n'=N'/(2L_c)$ is a discontinuous function of $\mu$ which would give rise to discontinuous behavior of the self-energy and thus of all observables.
This fact elucidates the second major motivation for introducing one bath site per correlated site:
Bath sites serve as charge reservoirs.
The entries $i,j$ of the self-energy $\Sigma_{ij\sigma}(\omega)$ are restricted to the {\em correlated} sites, and for a half-filled cluster, i.e.\ $n'=1$ or $N'=2L_c$, the electron density on the correlated sites can vary in the entire range from $n'_c=0$ to $n'_c=2$ (see Ref. \onlinecite{BHP08} for a detailed discussion). 
This is important not only for studies of density dependencies but also for ferromagnetic phases. 
With the help of the bath sites, an arbitrary and continuous variation of the cluster magnetization $m'=n'_{c\uparrow}-n'_{c\downarrow}$ can be achieved in the same way. 
We will comment on this in the discussion of the results below.

\section{Numerical evaluation}
\label{NUM}

The actual calculations are performed for finite Hubbard chains (Hamiltonian $H$, \refeq{hubbard}) with typically $L = \mathcal{O}(10^3)$ sites and assuming periodic boundary conditions.
It is convenient to consider the reference system (Hamiltonian $H'$) as being composed of $N_{k}=L/L_c$ identical and 
disconnected clusters consisting of $L_c$ sites each, i.e.\ the sites of the reference system form a translationally invariant superlattice of $N_k$ supersites.
Via the self-energy, this superlattice structure is also imposed on the expectation values of observables of the original model.

The grand potential at zero temperature can be calculated as: \cite{Pot03b}
\begin{equation}
\Omega(\ff \lambda) 
= 
\Omega' 
- \sum_{k,n} \omega'_n \, \Theta(-\omega'_n)
+ \sum_{k,n} \omega_n(k)\, \Theta(-\omega_n(k)) \: .
\labeq{poles}
\end{equation}
Here, $\Omega'$ is the grand potential of the reference system and $\omega'_n$ are the poles of the one-electron Green's function of the reference system. 
These can be calculated exactly by means of the Lanczos approach. \cite{LG93}
Note that the $\omega'_n$ do not depend on the wave ``vectors'' $k$ of the first Brillouin zone of the superlattice
as the clusters are disconnected, and thus the $k$-sum in the second term simply yields a factor $N_k$.
$\omega_n(k)$ are the poles of the (approximate) Green's function of the model \refeq{hubbard}.
Finally, $\Theta$ denotes the Heaviside function, and the $\ff\lambda$ dependence of $\Omega'$ and of the poles $\omega'_n$ and $\omega_n(k)$ is implicit. 

There are several technical points which are essential for a reliable numerical evaluation of the VCA and therewith for the interpretation of the results. 
One of the major intentions of the present paper is to show how one can efficiently deal with the different finite-size effects in the evaluation of \refeq{poles}, in particular close to a second-order phase transition, and with the problem of finding stationary points in a high-dimensional parameters space.

\subsection{Exact frequency summation}

The first problem consists in the infinite sums over Matsubara frequencies in \refeq{poles}.
For not too large reference systems, this is performed conveniently and numerically exact by means of the so-called Q-matrix technique: \cite{AAPH06b,BHP08} 
Let $\ff \Lambda'$ be the diagonal matrix with the poles $\omega'_n$ of the cluster Green's function $\ff G'$ as
diagonal elements.
The Lehmann representation of $\ff G'$ can then be written as 
\be
\ff G'(\omega) = \ff Q \frac{1}{\omega - \ff \Lambda'} \ff Q^\dagger \: ,
\ee
with an appropriate weight matrix $Q_{(i\sigma),n}$. 
Note that $\ff Q \ff Q^\dagger = \ff 1 \ne \ff Q^\dagger \ff Q$.
Using this Lehmann representation in the definition $\ff G_k = (\ff G_{0,k}^{-1} - \ff \Sigma)^{-1}$ of the Green's function of the original model, it is easy to see that the poles $\omega_n(k)$ of $\ff G_k$ then can be obtained as
eigenvalues of the matrix
\be 
\ff M(k) = \ff \Lambda' + \ff Q^\dagger \ff V(k) \ff Q \; .
\ee
Here $\ff V(k) = \ff \varepsilon(k) - \ff t'$ is the difference between the one-particle parameters of the original and
the reference system where the matrix $\ff\varepsilon(k)$ is the Fourier transform of $\ff t$ with respect to the superlattice.
In practice, the efficiency of the $Q$-matrix technique is set by the dimension of $\ff M(k)$, i.e.\ the number of poles of $\ff G'$, which, using Lanczos as a cluster solver, typically amounts to ${\cal O}(100)$.
For $L_c < 8$, the repeated diagonalization of $\ff M(k)$ for all $k$, and for $L_c>8$ the Lanczos diagonalization of $H'$ represents the dominant contribution to the necessary total CPU time, respectively.

\subsection{Interpolative $k$-summation}

The $k$-summation is much more tedious.
From \refeq{poles} it can be read off that $\Omega(\ff \lambda)$ is a non-analytic function for finite $L$. 
We first discuss the (approximate) one-electron excitation energies $\omega_n(k)$ of the original system.
A sign change of one of the energies $\omega_n(k)$ as function of a variational parameter $\lambda_i$ causes a kink in $\Omega(\ff \lambda)$ due to the $\Theta$ function.
Such kinks have a negligible relative weight in the $k$-sum and can be ignored in the thermodynamic limit $L\to\infty$ (if the interacting density of states at the Fermi edge stays finite).
For finite $L$ and in regions of the parameter space where $\Omega(\ff \lambda)$ is nearly flat, however, the mentioned kinks may lead to artifacts or at least to severe convergence problems for numerical techniques to find stationary points, particularly if derivatives of $\Omega(\ff \lambda)$ are required.

In principle, this finite-size effect can be controlled by increasing the system size $L$ and thereby the number of $k$-points $N_k$.
Although the computational effort is only linear in $N_k$, we found it to be much more effective to employ an interpolation algorithm which artificially increases the number of $k$-vectors while keeping the system size fixed. 
For two adjacent $k$-vectors $k_1$ and $k_2$, we interpolate between the excitation energies $\omega_n(k_1)$ and $\omega_n(k_2)$ instead of calculating $\omega_n(k)$ for intermediate $k$ by diagonalization of $\ff M(k)$.
Simple linear interpolation turns out to be sufficient.
The effect is a smoothing of the function $\Omega(\ff \lambda)$ which considerably stabilizes the subsequent optimization procedure without a significant increase of the computational effort.

Which pairs of poles at $k_1$ and $k_2$ correspond to each other, respectively, is actually unknown (as long as one does not analyze the corresponding eigenvectors of $\ff M(k)$) but for practical purposes it is sufficient to sort the respective pole sets and assume the $n$-th pole at $k_1$ to correspond to the $n$-th pole at $k_2$, i.e.\ ``level'' crossing is excluded.
If the number of $k$-points is sufficiently large, the interpolation procedure affects contributions to the sum over $k$ and $n$ in the last term of \refeq{poles} only in those cases where $\omega_n(k)$ crosses zero. 
Simple continuity arguments then show that a ``level'' crossing is unlikely in those cases, i.e.\ the possible error of disregarding ``level'' crossings is ${\cal O}(1/N_k)$.
This simple idea can also easily be generalized to higher dimensions.

\begin{figure}
\centerline{\includegraphics[width=.99\columnwidth,clip=]{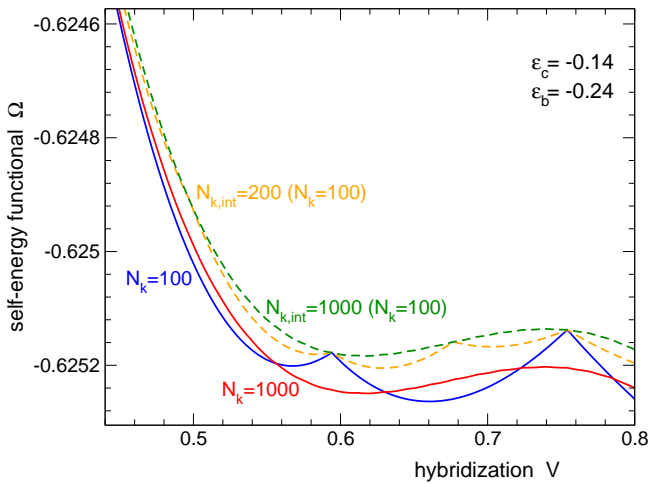}}
\caption{
Self-energy functional $\Omega(\ff \Sigma(V,\varepsilon_c,\varepsilon_b)) \equiv \Omega(V,\varepsilon_c,\varepsilon_b)$, plotted versus the hybridization strength $V$ while keeping the on-site energies fixed at their optimal values, $\varepsilon_c = -0.14$ and $\varepsilon_b = -0.24$.
Calculation for $t_1=1$ (this sets the energy scale throughout the paper), $t_2=0$, $U=4$, $\mu=-0.3$ and using the $L_c=2$ reference system.
Solid lines represent calculations without the interpolation method, dashed lines refer to calculations with a total (original plus interpolated) number of poles $N_{k,int}$.
}
\labfig{interpol}
\end{figure}

As an example to illustrate the interpolation scheme we show in \reffig{interpol} the self-energy functional as function of the hybridization $V$ for a reference system with $L_c=2$.
Convergence on the scale of the figure is obtained with $N_k \approx 1000$ (red solid line). 
As can be seen, a comparatively smooth curve can also be obtained with $N_k=100$ original but additional 900 interpolated $k$-points (dashed green line).
The comparison shows that the trend of $\Omega(V)$ is essentially unaffected by the interpolation scheme.
This merely produces a tiny overall shift of $\Omega(V)$ which is irrelevant for the determination of minima and maxima.
Using much less (yellow dashed) or no interpolated $k$-points (blue solid line) introduces the above-mentioned kinks.
For the example shown here, where the SFT grand potential is rather flat, these kinks would render a reliable determination of the optimal hybridization strength impossible.

\subsection{Level crossing in the reference cluster}
\label{sec:lc}

Let us now turn to the second possible source of a non-analytic behavior of $\Omega(\ff \lambda)$, namely a sign change of one of the single-electron excitation energies $\omega'_n$ of the reference system as a function of a variational parameter.
Consider an electron-removal process, for example.
Here $\omega'_n = E_0(N') - E_n(N'-1) - \mu \le 0$ where $E_n(N')$ is the $n$-th excited eigenenergy in the invariant subspace of $H'$ with (total) particle number $N'$.
If, as a function of $\ff \lambda$, the excitation energy $\omega'_n \to 0$, the ground state of the reference system becomes degenerate with an eventually new ground state in the $N'-1$ subspace.
Hence, $\omega'_n = 0$ would indicate a level crossing and a {\em discontinuous} change of the ground state of the reference system.
This in turn would induce a discontinuous change of the self-energy and thus a discontinuity of the SFT grand potential $\Omega(\ff \lambda)$ which was unphysical for obvious thermodynamical reasons.
It is therefore of utmost importance to keep $N'=\mbox{const}$, i.e.\ to ensure that the stationary point of the SFT functional (and a finite environment in parameter space) always corresponds to the same $N'$. 
The same holds for $z$-component of the {\em total} spin.

In principle, one might try to ignore a sign change of $\omega'_n$ as a function of the {\em optimal} $\ff \lambda$, i.e.\ as a function of a model parameter, and formally calculate the self-energy from the ground state of a subspace with {\em given} $N'$. 
Besides the fundamental problem that this would actually correspond to a non-equilibrium situation, such a procedure also cannot work in practice: 
As the above discussion has shown, $\omega'_n = 0$ would then induce a strong kink in $\Omega(\ff \lambda)$ which cannot be smoothed unless extremely large reference clusters with $L_c \to \infty$ are considered.

\subsection{Local optimization}

Tracing a stationary point of $\Omega(\ff \lambda)$ as a function of a model parameter can be accomplished by a {\em local} technique, i.e.\ assuming the stationary point $\ff \lambda_{\rm st}$ to be close to a starting point $\ff \lambda_0$.
A naive application of Newton's method to find a zero of $\nabla \Omega(\ff \lambda)$ has turned out to be inefficient, however, since equipotential surfaces $\Omega(\ff \lambda) = \mbox{const.}$ are usually highly anisotropic. 
Below we briefly describe our modified algorithm which uses an adaptive local coordinate frame with a directionally dependent calculation of partial derivatives.

Let $\ff \lambda_n$ denote the set of variational parameters at iteration step $n$.
Assuming $\Omega(\ff \lambda)$ to be approximately given by a quadratic form, the next estimate is
\begin{equation}
\ff \lambda_{n+1} = \ff \lambda_{n} - \ff H^{-1}_n  \left[\ff
\nabla \Omega(\ff \lambda) 
\right]_{\ff \lambda = \ff \lambda_n} \; ,
\end{equation}
where
\begin{equation}
H_{n,ij} = \left. 
\frac{\partial^2 \Omega(\ff \lambda)}{\partial \lambda_i \partial \lambda_j}
\right|_{\ff \lambda = \ff \lambda_n} 
\end{equation}
is the Hessian at $\ff \lambda_n$.
A numerically stable evaluation of the Hessian (and the gradient) is crucial here. 
This can be achieved iteratively by principal axis transformation:
\be
\ff H_n = \ff U_n \ff D_n \ff U_n^{\rm T} \: .
\ee
The diagonal matrix $\ff D_n$ contains the eigenvalues of $\ff H_n$.
New coordinates $\widetilde{\ff \lambda}$ are defined via the orthogonal transformation
\be
\widetilde{\ff \lambda} = \ff U_{n}^{\rm T} \ff \lambda \: .
\ee
The Hessian for the next $n+1$-st step is calculated in the new frame:
\be
\widetilde{H}_{n+1,ij} = \left. 
\frac{\partial^2 \Omega(\widetilde{\ff \lambda})}
{\partial \widetilde{\lambda}_i \partial \widetilde{\lambda}_j} 
\right|_{\widetilde{\ff \lambda} = \widetilde{\ff \lambda}_{n+1}} \: .
\ee
Here $\Omega(\widetilde{\ff \lambda}) \equiv \Omega(\ff \lambda(\widetilde{\ff \lambda}))$. 
Inverse transformation
\be
\ff H_{n+1} = \ff U_n \widetilde{\ff H}_{n+1} \ff U_n^{\rm T}	
\ee
yields the new Hessian in the original frame which is required for the next iteration step.
The main point is that in principal coordinates we have
\be
\Omega(\widetilde{\ff \lambda}) - \Omega(\widetilde{\ff \lambda}_{\rm st}) = 
\frac{1}{2} \sum_i \left.\frac{\partial^2 \Omega(\widetilde{\ff \lambda})}
{\partial \widetilde{\lambda}_i^2}\right|_{\widetilde{\ff \lambda} = \widetilde{\ff
\lambda}_{\rm st}} (\widetilde{\lambda}_i - \widetilde{\lambda}_{{\rm st},i})^2
\labeq{taylorst}
\ee
for $\ff \lambda$ close to $\ff \lambda_{\rm st}$. 
Hence, $\widetilde{\ff H}_{n+1}$ becomes almost diagonal and can be calculated as a difference quotient using discrete steps which depend on the principal direction:
\be
\Delta\widetilde{\lambda}_i = \sqrt{\Delta \Omega/ \widetilde{H}_{n,ii}} \: .
\ee
Here $\Delta\Omega$ is a suitably chosen constant.
This implies that partial derivatives along directions in parameter space where $\Omega(\ff \lambda)$ is almost flat are computed with a large $\Delta \widetilde{\lambda}_i$, while a small $\Delta \widetilde{\lambda}_i$ is used along directions where $\Omega(\ff \lambda)$ is strongly curved.
This has turned out to be crucial for a numerically stable algorithm.

\subsection{Global optimization}

As a prerequisite for a local method to trace a stationary point, a global method must be available which is applicable even if a reasonable starting point is not known.
Except for global minimization algorithms which can be applied to minimize $|\nabla \Omega(\ff \lambda)|^2$, for example, there is no general global technique to find stationary points in a multidimensional space unfortunately.

\begin{figure}
\centerline{\includegraphics[scale=.26,clip=]{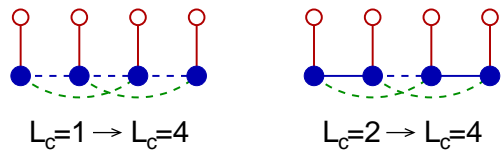}}
\caption{
Crossover from the $L_c=1$ to the $L_c=4$ reference system (left) and from the $L_c=2$ reference system to the one with $L_c=4$ (right).
Dashed lines represent intra-cluster hopping parameters scaled by a factor $\alpha$.
It is assumed that a stationary point of the SFT functional is known for the case where the intra-cluster hopping parameters are switched off ($\alpha=0$). 
The stationary point is traced {\em locally} while adiabatically switching on the hopping parameters, $0 < \alpha < 1$. 
This yields a stationary point for the respective $L_c>1$ reference system ($\alpha=1$).
}
\labfig{crossover_refsys}
\end{figure}

In context of the SFT, however, there is an elegant solution to this problem since any local method can be converted into a global one with the help of a crossover procedure as has been pointed out in \ocite{Ede08}.
The main idea is to modify the {\em original} system by switching off the intercluster hopping (in the same way as it is done in the reference system). 
For this truncated system, the VCA trivially yields the exact solution, and the stationary point is trivially given by one-particle parameters of the reference system which are equal to those of the truncated one.
One then adiabatically switches on again the inter-cluster hopping in the truncated system, i.e.\ one replaces $\ff t_{\rm inter} \to \alpha \ff t_{\rm inter}$ and increases the parameter $\alpha$ from $\alpha=0$ to $\alpha=1$.
During this adiabatic process the stationary point of the reference system $\ff \lambda_{\rm st}(\alpha)$ can be traced by means of a {\em local} optimization method.
Finally, $\ff \lambda_{\rm st} = \ff \lambda_{\rm st}(\alpha=1)$ is the stationary point of the original system.

Here, we present a variant of this crossover trick which makes use of the fact that it has turned out to be easy to globally find a stationary solution for the $L_c=1$ reference system (i.e.\ for the dynamical impurity approximation).
An adiabatic crossover from the $L_c=1$ reference system to an $L_c>1$ reference system can be performed then by introducing a dimensionless parameter $\alpha$ to scale the nearest-neighbor and next-nearest-neighbor hopping in the {\em reference} system: $t'_1 \to \alpha t'_1$ and $t'_2 \to \alpha t'_2$.
For $\alpha=0$ we recover the mean-field solution, for $\alpha=1$ we have the VCA with $L_c>1$. 
This procedure can be applied to cross over between two arbitrary reference systems.
Examples are given in \reffig{crossover}.

For our calculations using the $L_c=4$ reference system, we started from $L_c=1$ and have changed $\alpha$ from 0 to 1 in steps of 0.05.
An example is shown in \reffig{crossover}. 
The crossover procedure can be done along two different routes, namely from $L_c=1$ to $L_c=4$ directly and, in two crossover steps, from $L_c=1$ via $L_c=2$ to $L_c=4$.
The resulting optimal variational parameters are the same for both routes as can be seen from the figure.
It is physically plausible that with increasing $\alpha$ and cluster size, the values of the optimal parameters
tend to approach the ``physical values'' of the original system, i.e.\ $V=0$ and $\varepsilon_c=0$ while for $L_c=1$ stronger deviations are necessary to partially compensate for the effect of the truncated inter-cluster hopping. 

\begin{figure}
\centerline{\includegraphics[width=0.99\columnwidth,clip=]{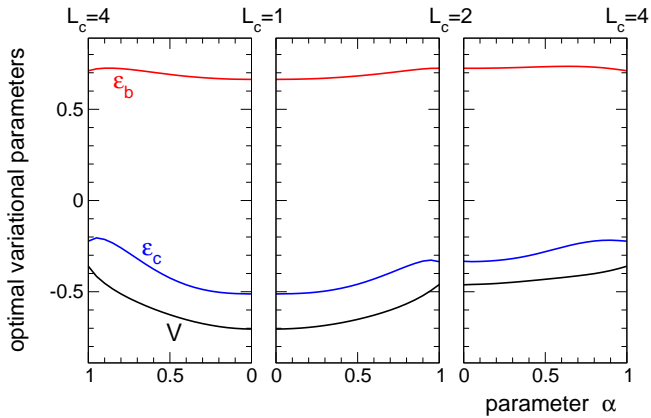}}
\caption{Optimal variational parameters $\varepsilon_c$, $\varepsilon_b$
and $V$ during the crossover procedure from $L_c=1$ to $L_c=4$ (left panel, from right to left) and from 
$L_c=1$ to $L_c=4$ via $L_c=2$ (middle and right panel).
Both crossover procedures yield the same optimal parameters at $\alpha=1$.
Calculations for an original system with $L=4000$ sites, $t_1=1$, $t_2=-0.2$, $U=4$ and $\mu=0.7$. 
}
\labfig{crossover}
\end{figure}

\subsection{Calculations at fixed density and magnetization}

Self-energy-functional theory has originally been developed using the grand-canonical ensemble. 
At zero temperature, starting from a grand-canonical Hamiltonian with chemical potential $\mu$ and magnetic field $B$,
\begin{equation}
\mathcal{H} = H - \mu \sum_{i} (n_{i\uparrow} + n_{i\downarrow}) - B \sum_i (n_{i\uparrow} - n_{i\downarrow})  \: ,
\end{equation}
the SFT grand potential $\Omega=\Omega(\ff \lambda,\mu,B)$ is a function of the variational parameters $\ff \lambda$ [see \refeq{omegalambda}] and of $\mu$ and $B$ (and other model parameters, such as $U$). 
The variational parameters are fixed by $\partial \Omega(\ff \lambda,\mu,B) / \partial \ff \lambda = 0$ while, at the respective stationary point, the derivatives with respect to $\mu$ and $B$ yield the expectation values of the total particle number and magnetic moment:
\be
\langle N \rangle = - \frac{\partial \Omega(\ff \lambda,\mu,B)}{\partial \mu}  \; , \;
\langle M \rangle = - \frac{\partial \Omega(\ff \lambda,\mu,B)}{\partial B} \: .
\ee
We also define the electron density $n=N/L=\sum_{i\sigma}\langle n_{i\sigma} \rangle/L$ and the 
magnetization $m=M/L=\sum_{i\sigma}z_\sigma\langle n_{i\sigma} \rangle/L$ with $z_{\uparrow,\downarrow} = \pm 1$.

To construct phase diagrams it is much more convenient, however, to keep $n$ instead of $\mu$ fixed, e.g.\ to study $U$ dependencies at fixed density $n$.
If there is no manifest particle-hole symmetry (off half-filling or for $t_2\ne 0$), the corresponding chemical potential is not known a priori. 
Furthermore, it is highly desirable to perform calculations at fixed $m$ instead of $B$ to search for a ferromagnetic phase: 
Starting from a paramagnetic solution with $m=0$ and adiabatically increasing $m$, one simply has to trace the solution and find the corresponding $B=B(m)$.
A spontaneous ferromagnetic solution is then indicated by a finite $m$ with $B(m)=0$.

Consider the twofold Legendre transformation from the grand potential $\Omega$ via the free energy $F=\Omega + \mu N$ to the Gibbs energy $G = \Omega + \mu N + B M$:
\be
\Omega(\ff \lambda,\mu,B) \mapsto F(\ff\lambda,N,B) \mapsto G(\ff \lambda,N,M) \: .
\labeq{legendre}
\ee
At given $N$ and $M$, the SFT Gibbs energy $G(\ff \lambda,N,M)$ is obtained from $G(\ff \lambda, \mu, B, N, M) \equiv \Omega(\ff \lambda,\mu,B) + \mu N  + B M$ via the original stationarity condition
\be
\frac{\partial G}{\partial \ff\lambda} = 0 
\Leftrightarrow
\frac{\partial \Omega}{\partial \ff\lambda} = 0 
\ee
and two additional conditions fixing $\mu$ and $B$
\ba
\frac{\partial G}{\partial \mu} = 0 
& \Leftrightarrow & 
\left\langle\sum_{i\sigma} n_{i\sigma} \right\rangle 
= 
N
\\
\frac{\partial G}{\partial B} = 0 
& \Leftrightarrow & 
\left\langle\sum_{i\sigma} z_\sigma n_{i\sigma} \right\rangle 
= 
M \: .
\ea
Hence, one simply has to consider $\mu$ and $B$ in addition to $\ff \lambda$ as variational parameters.

\begin{figure}
\centerline{\includegraphics[width=.9\columnwidth,clip=]{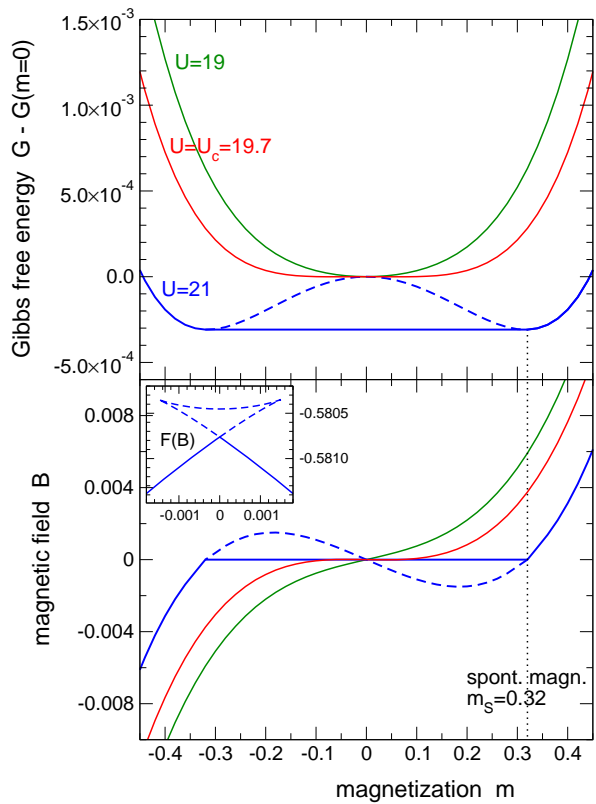}}
\caption{
Gibbs free energy $G$ (per site) and the external magnetic field $B$ as functions of the magnetization $m$.
Calculations have been performed at fixed density $n=0.7$ for $t_1=1$, $t_2 = -0.2$ and different $U$ (as indicated) using the $L_c=1$ reference system ($L=4000$).
For $U=21$ the dashed line shows the result of the actual calculation.
This solution, however, is unstable for $|m|<0.32$, and a solution with lower $G$ is obtained by a Maxwell construction (solid line).
The inset shows the free energy $F$ (per site) as function of $B$ for $U=21$. 
}
\labfig{B_M}
\end{figure}

\begin{figure}
\centerline{\includegraphics[width=0.9\columnwidth,clip=]{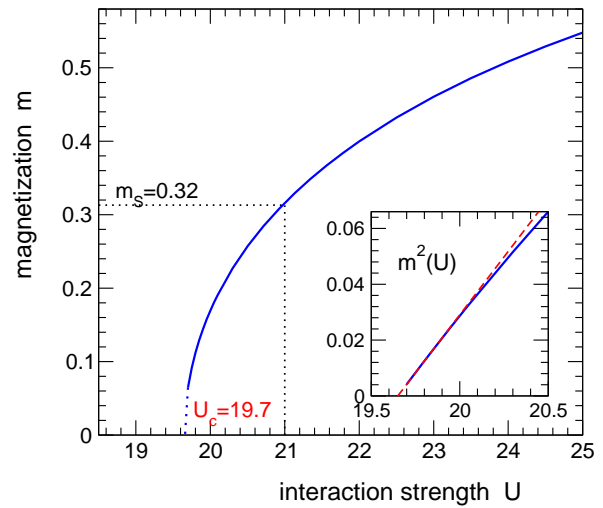}}
\caption{
Magnetization $m$ as a function of the interaction strength $U$.
For $U=21$ a spontaneous magnetization $m_s=0.32$ is obtained corresponding to the blue lines in \reffig{B_M}. 
Inset: $m^2$ as a function of $U$ close to $U_c$. 
The red dashed line is a linear extrapolation to $m=0$.
Parameters of the calculation: see \reffig{B_M}.
}
\labfig{M_U}
\end{figure}

To illustrate the method we have performed VCA calculations using the $L_c=1$ reference system at fixed density $n=0.7$ and different $U$ close to a second-order phase transition, see \reffig{B_M}.
The magnetization is treated as a given quantity.
The top panel shows the Gibbs free energy as function of $m$. 
For $U=19$ this is a convex function as it is prescribed by thermodynamic stability. 
$U_c=19.7$ marks a critical point above which the Gibbs energy becomes thermodynamically unstable within a certain range of magnetizations.
As can be seen in the top panel, the phase is locally unstable for $-0.18 < m < 0.18$ where the Gibbs energy is concave.
A thermodynamically stable state is obtained via a Maxwell construction. 
This yields the solid line.
Between $-0.32 < m < 0.32$ the Gibbs energy is a constant which implies $B = \partial G / \partial M = 0$. 
Hence, an infinitesimal field $B=0^+$ will produce a finite magnetization $m=0.32$.
States with $|m|<0.32$ can realized by macroscopy phase separation.
We conclude that $U_c=19.7$ marks a continuous transition from the paramagnetic state to a state with spontaneous ferromagnetic order.
The function $B(m)$ (lower panel in \reffig{B_M}) can be discussed analogously. 
Local thermodynamic instability is indicated by a negative slope, and instability with respect to a Maxwell constructed state is indicated by the dashed line.
Finally, the same physics can be seen by looking at the free energy (at $T=0$ equal to the ground-state energy $E = \langle H \rangle - B M$) given as a function of $B$ (see inset for $U=21$).
Note that $m$ can be computed as a derivative of $G$ or via an integration of the spin-dependent local density of states which, due to the thermodynamical consistency of the SFT, yields the same result.

It is interesting that the mean-field ($L_c=1$) approach yields a critical interaction $U_c=19.7$ which is rather close to the numerically exact result $U_c=18.5$ obtained via density-matrix renormalization group. \cite{DN98}
Characteristic for a mean-field approach is the square-root behavior of the order parameter $m(U)$ close to the critical point, $m \propto \sqrt{U-U_c}$.
This can be seen in \reffig{M_U}, where the magnetization is displayed as function of $U$ for fixed density $n=0.7$.
The inset shows a linear trend of $m^2$ close to $U_c$.

\subsection{Ferromagnetic susceptibility}

\begin{figure}
\centerline{\includegraphics[width=.99\columnwidth,clip=]{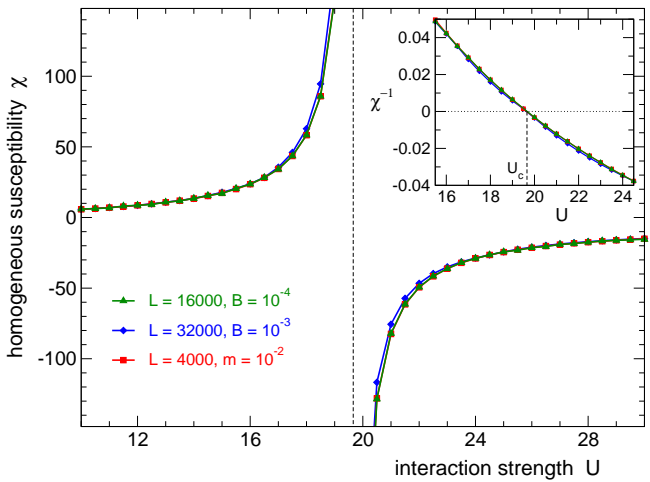}}
\caption{
Homogeneous static magnetic susceptibility $\chi = \partial m / \partial B|_{B=0}$ (per site) as a function of the interaction
strength $U$ using the $L_c=1$ reference system ($t_1=1$, $t_2 = -0.2$, $n=0.7$).
The inset shows $\chi^{-1}(U)$.
Results are obtained by three different techniques.
Green lines: spin-dependent optimization of the variational parameters at fixed small field $B$.
Blue lines: spin-independent parameter optimization according to \refeq{chi_eder}.
Red lines: optimizing the Gibbs energy by varying the external field at fixed small magnetization.
Convergence is obtained for different respective system sizes $L$ as indicated.
}
\labfig{chi}
\end{figure}

For the calculations of $m(U)$, as displayed in \reffig{M_U}, the simultaneous optimization of 8 variational parameters is required, namely the spin-dependent on-site energies $\varepsilon_c$ and $\varepsilon_b$, the spin-dependent hybridization strength $V$ plus $\mu$ and $B$.
The number of parameters can be reduced if one is interested in the phase boundaries only.
Rather than tracing a spontaneously symmetry-broken solution, a second-order critical point can be found from the divergence of a suitably defined susceptibility. 
Here we consider the homogeneous static magnetic susceptibility $\chi = -(1/L)\; \partial^2 F / \partial B^2$. 
The most obvious way to calculate $\chi$ is to apply a small external homogeneous magnetic field $B$ and to look at the linear response $m$, i.e.\ $\chi = \lim_{B\to 0} m / B$.
For this case, it is convenient to consider $B$ as fixed which implies that only 7 variational parameters (for $L_c=1$ or $L_c=2$) have to be taken into account.
The result for the $L_c=1$ reference system is shown in \reffig{chi} (green lines). 
The divergence of $\chi$ at $U_c=19.7$ is consistent with the $U_c$ extracted from the order parameter in \reffig{M_U}.
As it is typical for a mean-field approach $\chi^{-1}$ (see inset) is a linear function of $U$ close to $U_c$.
The same holds for $L_c=2$. 
Again this has to be expected as critical phenomena should not depend on the reference cluster size.

For the calculation of the susceptibility, a variational optimization of {\em spin-dependent} variational parameters is actually not necessary as was recognized by Eder. \cite{Ede09}
This can be seen in the following way.
Consider the free energy $F=F(\ff \lambda,B)$ [\refeq{legendre}] where we suppress the $N$ dependence in the notation. 
Due to the stationarity conditions, $\partial F(\ff \lambda,B) / \partial \ff \lambda = 0$, the optimal $\ff \lambda$ can be considered as a function of $B$, i.e.\ $\ff \lambda = \ff \lambda(B)$. 
Therefore, 
\be
\frac{d}{dB} \frac{\partial F}{\partial \ff \lambda}(\ff \lambda(B),B) = 0 \: .
\labeq{cond}
\ee
Carrying out the differentiation, we find
\be
\frac{\partial^2 F}{\partial \ff \lambda \partial \ff \lambda}(\ff \lambda(B),B)
\frac{d \ff \lambda(B)}{dB} 
+
\frac{\partial^2 F}{\partial B \partial \ff \lambda}(\ff \lambda(B),B)
= 0 \: .
\labeq{cond1}
\ee
This is a linear set of equations which can be solved by matrix inversion to get
\be
\frac{d \ff \lambda(B)}{dB} 
= -
\left[\frac{\partial^2 F}{\partial \ff \lambda \partial \ff \lambda}\right]^{-1}
\frac{\partial^2 F}{\partial B \partial \ff \lambda}
\: .
\labeq{cond2}
\ee
Now, the susceptibility is given by $\chi = - d^2 F(\ff \lambda(B),B) / dB^2$. 
Hence
\be
\chi = - \frac{d}{dB} \left( \frac{\partial F(\ff \lambda(B),B)}{\partial \ff \lambda} \frac{d \ff \lambda(B)}{dB} + \frac{\partial F(\ff \lambda(B),B)}{\partial B} \right) \: .
\ee
Using \refeq{cond} and the stationarity condition once more, we see that the first term does not contribute, and thus
\be
\chi
= 
- \frac{\partial^2 F(\ff \lambda,B)}{ \partial \ff \lambda \partial B} \frac{d \ff \lambda(B)}{dB}
- \frac{\partial^2 F(\ff \lambda,B)}{\partial B^2} \: ,
\labeq{chi_eder}
\ee
where $d \ff \lambda(B)/d B$ can be eliminated using \refeq{cond2}.
Consequently, for the calculation of $\chi$ it is sufficient to consider a paramagnetic state and to optimize spin-independent variational parameters only.
This strongly reduces the computational effort.
Once a paramagnetic stationary point is found, partial derivatives according to \refeq{chi_eder} and \refeq{cond2} have to calculated with spin-dependent parameters $\ff \lambda$ in a final step.
The resulting $\chi$ as a function of $U$ is shown in \reffig{chi} as the blue lines. 

A third way to determine the susceptibility is to keep the magnetization $m$ fixed at a small value and vary the field $B$ to optimize the Gibbs energy $G$, see red line in \reffig{chi}.
Here, a divergence of $\chi$ is indicated by $B(m \neq 0) = 0$.
This calculation involves spin-dependent parameter optimization and to take the field $B$ as a variational parameter in addition.

Finite size effects play a crucial role for the calculation of the susceptibility in the critical regime.
It turns out that the numerically most expensive calculation where the magnetization is kept fixed is most stable against finite size effects. 
On the scale of \reffig{chi} converged results are obtained for a comparatively moderate system size of $L=4000$ sites.

\section{Results and discussion}
\label{FM}

\subsection{Ferromagnetism in the Hubbard model}

Applying the Hartree-Fock approximation to the single-band Hubbard model, one is lead to the Stoner criterion, \cite{Sto81}
\be
  U \rho_0(0) > 1 \; ,
\ee
for the existence of a ferromagnetic instability.
Therewith, the calculation of the free ($U=0$) local density of states (DOS) $\rho_0(\omega)$ at the Fermi edge $\omega=0$ can give first insights where ferromagnetism is likely to occur. 
Conceptually, however, the Hartree-Fock approach is a static mean-field theory, and quantum fluctuations are neglected altogether. 
If at all, reliable results can be derived for the extreme weak-coupling regime where ferromagnetism is unlikely to occur.

Despite the simplicity of the Hubbard model, only a few rigorous results on ferromagnetism are available. \cite{Lie94,Lie95,Tas96}
The Mermin-Wagner theorem \cite{MW66,Gho71} excludes spontaneous breaking of the SU(2) symmetry for finite temperatures and dimensions lower then three.
For the one-dimensional case and nearest-neighbor hopping, Lieb and Mattis \cite{LM62} have shown that the ground state for any even number of electrons is always a non-magnetic singlet independent of $U$.
A ferromagnetic ground state is also excluded in the low-density limit $n \mapsto 0$ irrespective of $U$ as has been argued by Kanamori. \cite{Kan63}
His $T$-matrix approach, however, must be based on the assumption that weak-coupling perturbation expansion converges.
Lieb \cite{Lie89} has shown that a ferromagnetic ground state is excluded for a bipartite lattice with nearest-neighbor hopping at half-filling $n=1$ any $U>0$ independent of the dimensionality. 
As has been demonstrated by Nagaoka \cite{Nag66} the ground state of the half-filled model with one hole added is fully polarized for $U=\infty$ on bipartite lattices (for fcc and hcp lattices with negative hopping integrals) in three or higher dimensions.
While criteria for the stability of the fully polarized state for thermodynamically relevant dopings could not be obtained, \cite{DW89,Tia91,BRY90} it is possible to reduce the parameter space left for a stable Nagaoka state in the thermodynamic limit by different variational approaches. \cite{WUMH96,HUMH97} 
Mielke and Tasaki \cite{Mie91,Tas92,MT93,Tas98} proved the stability of ferromagnetism for special lattices, such as the Kagom\'e lattice, for which there are dispersionless parts of the Bloch band (``flat-band ferromagnetism''). 
In these systems the Fermi sea is degenerate with ferromagnetic states for $U=0$, and ferromagnetism becomes stable for $U>0$.
M\"uller-Hartmann \cite{MH95} has considered the one-dimensional model with a next-nearest-neighbor hopping such that the free band has two degenerate minima. 
In the low-density limit a metallic ferromagnetic ground state is obtained due to ferromagnetic exchange in a corresponding effective two-band model.
Similarly, Tasaki \cite{Tas95} constructed a one-dimensional Hubbard model with next-nearest neighbor hopping which has a ferromagnetic (insulating) ground state at quarter filling and sufficiently strong $U$.

\subsection{Ferromagnetism in infinite dimensions}
\label{sec:infd}

A comprehensive but approximate approach to ferromagnetic order is provided by dynamical mean-field theory. \cite{GKKR96,PJF95,KV04}
In the past several studies have addressed the magnetic phase diagram of the Hubbard model on infinite-dimensional lattices where the DMFT becomes exact. 
At least two routes towards ferromagnetic order could be identified:
(i) On a particle-hole symmetric hypercubic lattice ferromagnetism is realized for very strong Coulomb interaction $U$ and fillings close to half-filling. \cite{OPK97,ZPB02,PHMK08}
This is reminiscent of the Nagaoka state. \cite{Nag66}
(ii) On the other hand, a moderate Hubbard-$U$ is sufficient for lattices with a free DOS exhibiting a pronounced asymmetry. \cite{Ulm98,WBS+98,Uhr96} 
Here a ferromagnetic ground state is observed in large regions of the $U$-$n$ phase diagram. 
A simple mechanism for ferromagnetic order is not apparent although some understanding could be achieved \cite{HN97b,PHWN98,WBS+98} by techniques and arguments related to the Hubbard-I approach. \cite{Hub63}
The Stoner criterion turns out to be inadequate.
Furthermore, also a realization of flat-band ferromagnetism \cite{Tas98} can be found \cite{PP09}
on a Bethe lattice with infinite coordination.

We have performed calculations for different lattices, i.e.\ for different free DOS, respectively, using the self-energy-functional approach for a reference system with one correlated and one bath site only, i.e.\ $L_c=1$, see \reffig{refsys}.
This is referred to as the dynamical impurity approximation (DIA) in the following.
Let us recall that the DIA with an infinite number of bath sites would exactly correspond to DMFT. \cite{Pot03a}
While local quantum fluctuations are treated exactly within DMFT, the DIA is much simpler and includes some local fluctuations only.
However, due to the presence of the bath site it allows for the formation of a local (Kondo-type) singlet. 
Our goal is here to test the DIA by comparing with available DMFT results for the ferromagnetic phase.
This serves as a benchmark of the approximation.
Furthermore, as a computationally cheap method, the DIA allows for a more comprehensive study of the phase diagram.

\begin{figure}[t]
\includegraphics[width=0.85\columnwidth]{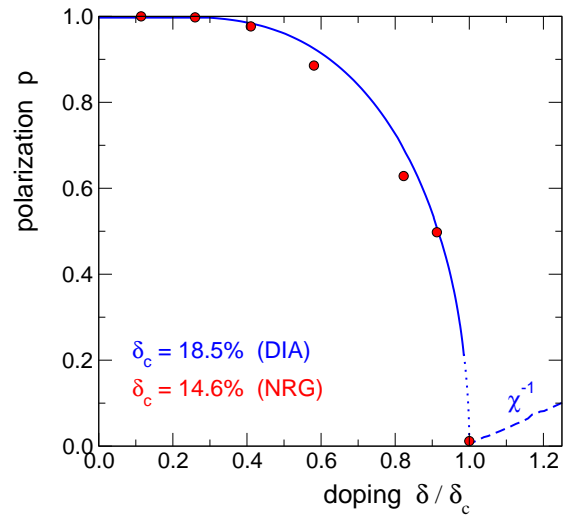}
\caption{Polarization $p = m/n$ (with $m=n_\uparrow - n_\downarrow$ and $n=n_\uparrow + n_\downarrow$) as a function of the doping $\delta = 1 - n$ for the hypercubic lattice in infinite dimensions at $U=50$. 
$\delta_c$ is the critical doping.
The energy scale is set by the variance $\sigma^2 = 0.5$ of the Gaussian free DOS.
Results of the DIA (lines) are compared with data from full DMFT-NRG calculations (points) taken from Zitzler et al.\ \cite{ZPB02}
Dashed line: inverse homogeneous static susceptibility $\chi^{-1}$ as obtained from the DIA.
}
\labfig{mag}
\end{figure}

We start with the first route (i) towards ferromagnetism and consider fillings close to half-filling and very strong Coulomb interaction. 
Here we can compare with DMFT results by Zitzler et al.\ \cite{ZPB02} which have been obtained using the numerical renormalization group (NRG) as an impurity solver.
The calculations have been carried out for the hypercubic lattice in infinite dimensions.
Using the same conventions as in Ref.\ \onlinecite{ZPB02}, the free DOS is given by 
\begin{equation}
  \rho_0(\omega) = \frac{1}{\sqrt{\pi}} e^{-\omega^2} \: .
\end{equation}
The variance of the Gaussian DOS $\sigma^2 = 0.5$ sets the energy scale.

The results are displayed in \reffig{mag}. 
For very strong $U$ the DMRG-NRG data predict an almost fully polarized ferromagnetic state at low dopings $\delta = 1-n$. 
Note that as a consequence of the tails of the free DOS, the ground state cannot be fully polarized in a strict sense. \cite{FMMH90}
However, this exponentially small scale cannot be expected to be visible in the data.
With increasing doping the system undergoes a continuous phase transition to the paramagnet at a critical doping $\delta_c=14.6\%$.
The result of the DIA agrees well with the full DMFT and likewise predicts a continuous transition from a fully polarized state to the paramagnetic phase. 
On the rescaled plot in \reffig{mag}, the agreement is even quantitative.
However, the critical doping for the phase transition ($\delta_c=0.185$) is significantly overestimated as compared to DMFT-NRG ($\delta_c=0.146$).
As this means a stronger tendency towards ferromagnetism, one may conclude that the DIA underestimates the effect of local quantum fluctuations.

The DIA results are consistent in themselves: 
The magnetization $m$ can be calculated via the spectral theorem from the spin-dependent one-electron spectral function, or as the derivative of the optimal grand potential with respect to an external magnetic field.
Both computations yield the same result as has been checked numerically and as is clear from the formalism. \cite{AAPH06a} 
We also checked numerically that the Luttinger sum rule is fulfilled. 
Within the DIA this must be respected \cite{OBP07} - in the paramagnetic but also in the ferromagnetic state.
In the spin-polarized metallic phase there are two Fermi surfaces with Fermi-surface volumes for $\sigma=\uparrow,\downarrow$
\begin{equation}
   V_{{\rm FS}, \sigma} = 
   \sum_{\bf k} \Theta(\mu - \varepsilon({\bf k}) - \Sigma_\sigma(0))
\end{equation}
where $\varepsilon({\bf k})$ is the Bloch band dispersion and $\Sigma_\sigma(\omega)$ the $\ff k$-independent self-energy.
The Luttinger theorem then reads as
\begin{equation}
   V_{{\rm FS}, \sigma} \stackrel{!}{=} \langle N_\sigma \rangle =
   L \int_{-\infty}^0 d\omega \: \rho_\sigma(\omega) \: .
\end{equation}
For a local and real self-energy, the interacting local DOS can be written as $\rho_\sigma(\omega)=\rho_0(\omega+\mu-\Sigma_\sigma(\omega))$, and the Luttinger sum rule reads
$\mu=\mu_{0\sigma} + \Sigma_\sigma(0)$. 
Here $\mu_{0\sigma}$ is a (spin-dependent) chemical potential of the non-interacting system such that the spin-dependent particle numbers are the same as for the interacting system.

\begin{figure}[t]
\includegraphics[width=.88\columnwidth]{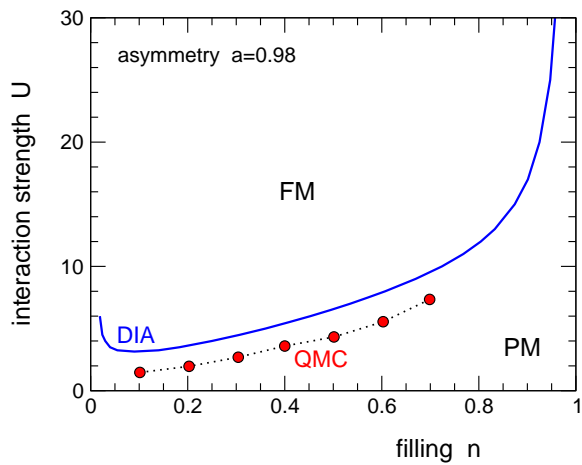}
\caption{
Ground-state phase diagram $U$ vs.\ filling $n$ for the asymmetric free DOS given by \refeq{asymDOS} with half band width $D=2$ and asymmetry parameter $a=0.98$.
The solid line refers to the DIA.
DMFT-QMC results (points) are taken from Wahle et al.\ \cite{WBS+98}
}
\labfig{qmcpd}
\end{figure}

\begin{figure}[t]
\includegraphics[width=.92\columnwidth]{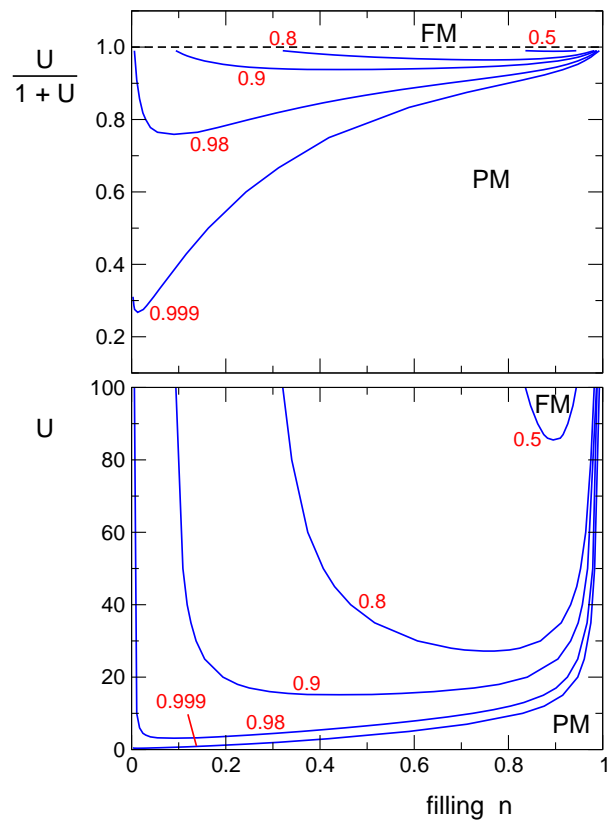}
\caption{
Phase diagram $U/(1+U)$ and $U$ vs.\ $n$, respectively, for various asymmetry parameters $a$. 
DIA calculations for the model free DOS with $D=2$ given in \refeq{asymDOS}.
}
\labfig{pdasy}
\end{figure}

Next we consider the second route (ii) towards ferromagnetism and consider a moderate Hubbard-$U$ but a free DOS with a pronounced asymmetry.
Here we can compare with the results of Wahle et al.\ \cite{WBS+98} who employed the Hirsch-Fye quantum Monte-Carlo method as an impurity solver for DMFT. 
Calculations have been performed for finite but low temperatures and could be extrapolated to extract a ground-state $U$-$n$ phase diagram which is shown in \reffig{qmcpd} (points).
The Bloch band dispersion $\varepsilon(\ff k)$ enters the DMFT (and also the DIA) via the free DOS only.
Hence, instead of specifying the lattice structure and the hopping parameters, one can likewise start from a certain model free DOS as input for the DMFT calculation. 
This has the advantage that the effect of the asymmetry of the free DOS can be studied systematically. 
In Ref.\ \onlinecite{WBS+98} the following model free DOS has been considered:
\be
  \rho_0(\omega) = c \frac{\sqrt{D^2-\omega^2}}{D + a \omega} \: .
\labeq{asymDOS}
\ee
Here, $a$ is a parameter which controls the asymmetry while the variance stays constant.
One can continuously tune the DOS from the symmetric case $a=0$, corresponding to the semielliptic DOS of the Bethe lattice with infinite coordination, over an asymmetric DOS with more and more spectral weight peaked in the vicinity of the lower band edge, to a DOS with an inverse square-root divergence at the band edge for $a=1$ eventually. 
Furthermore, in \refeq{asymDOS}, $c=(1+\sqrt{1-a^2})/(\pi D)$ is a normalization constant, and $D$ is the half band width which is set to $D=2$ to fix the energy scale.
The DMFT-QMC results in \reffig{qmcpd} correspond to a strongly asymmetric DOS characterized by $a=0.98$.

As is obvious from \reffig{qmcpd}, a ferromagnetic ground state is realized in large areas of the phase diagram. 
For low fillings a moderate Hubbard-$U$ is sufficient for ferromagnetism. 
With increasing $n$ the phase boundary $U_c(n)$ increases. 
Note that large $U$ values cannot be accessed easily within the Hirsch-Fye QMC approach.
It appears that this phase diagram is ruled by a mechanism that is completely different from the Nagaoka mechanism that has been suggested to rule the physics in case (i).
Contrary to the results shown in \reffig{mag}, ferromagnetism becomes {\em more} likely with increasing doping $\delta = 1-n$ and persists down to very small fillings.

\reffig{qmcpd} also shows the result of our DIA calculation for $a=0.98$ (solid line).
Again, we find a convincing qualitative agreement with the full DMFT. 
The phase boundary $U_c(n)$ shows the same trend but is systematically shifted towards higher interaction strengths.
We attribute this difference partly to the very sensitive dependence of the results on the asymmetry parameter.
This can be seen in \reffig{pdasy} where the result for the ground-state phase diagram from DIA calculations for different asymmetry parameters $a$ are given. 
It is obvious that for $a$ close to unity a tiny change of $a$ and thus of the free DOS results in a strong shift of the critical $U$.

Using the DIA one can easily trace the evolution of the phase diagram as a function of the asymmetry parameter. 
As can be see from \reffig{pdasy}, $U_c(n)$ can be very small for $a \to 1$, i.e.\ for the case where the free DOS diverges at the lower band edge.
For $a<1$ the critical interaction becomes large and eventually $U_c \to \infty$ for $n\to 0$.
With increasing $n$, however, the phase boundary soon develops a minimum at $n_{\rm min}$ and then becomes an increasing function of $n$.
This minimum is located at low fillings for asymmetry parameters close to unity but then shifts to higher fillings for a less asymmetric free DOS.
At the same time, $U_c(n_{\rm min})$ increases strongly. 
For $a=0.5$ we find $n_{\rm min} \approx 0.9$, and the ferromagnetic phase is confined to a small filling range close to half-filling and very strong Coulomb interaction.

It appears that the two routes towards ferromagnetism (Nagaoka vs.\ asymmetry of the free DOS) are linked continuously.
For even smaller asymmetry parameters $a<0.5$ ferromagnetism disappears completely. 
The symmetric case $a=0$ corresponds to a Bethe lattice with infinite coordination with a symmetric free DOS. 
Here a ferromagnetic state cannot be stabilized.
This is again consistent with full DMFT (NRG) calculations. \cite{PP09}
Obviously, the stability of the ferromagnetic state not only depends on the asymmetry $a$ but is also strongly affected by the detailed form of the {\em symmetric} free DOS since, as has been discussed above, for the symmetric Gaussian free DOS corresponding to the hypercubic lattice, there is again a ferromagnetic phase close to half-filling.
It is an open question whether the latter can really be attributed to the Nagaoka mechanism. 
On the one hand, the Nagaoka mechanism needs closed loops on the lattice which are present for the hypercubic one but absent for the Bethe lattice. 
On the other hand, the DMFT is sensitive to the lattice structure via the free DOS only.

Concluding, we find that the DIA gives qualitatively reliable results for the ferromagnetic ground-state phase diagram in all cases studied, as has been corroborated by the comparison with different full DMFT calculations.
A reference system with a single bath site appears to be sufficient to capture the main physics although quantitatively there is a tendency to overestimate the range where ferromagnetism is possible. 
While local quantum fluctuations are included in the DIA in a very simple way only, the approximation allows for the formation of a local (Kondo-type) singlet. 
Together with the internal thermodynamical consistency of the approach and with the fact that Luttinger's sum rule is respected, this ensures a reliable mean-field description of ferromagnetism.

\subsection{Ferromagnetism in one-dimensional chains}

\begin{figure}[b]
\centerline{\includegraphics[width=0.9\columnwidth,clip=]{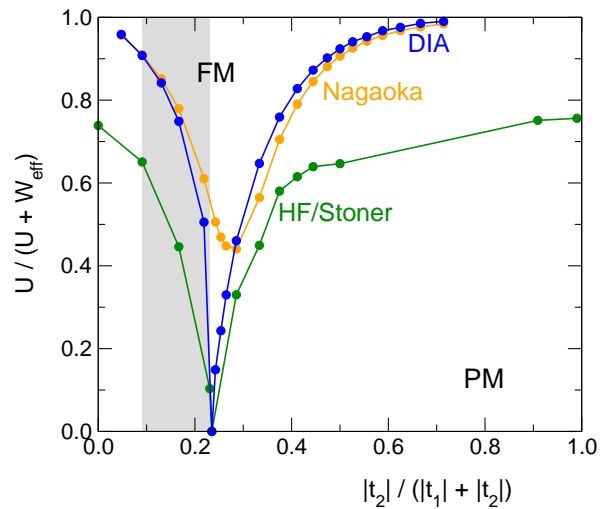}}
\caption{ 
Magnetic ground-state phase diagram $U$ vs.\ $t_2$ of the one-dimensional Hubbard model for quarter filling ($n=0.5$) as obtained by the dynamical impurity approximation (DIA, blue).
The shaded area represents the parameter regime for which a continuous transition is found.
$t_1=1$ fixes the energy unit. 
Only the range $t_2<0$ is considered.
Note the non-linear scales with $-\infty < t_2 < 0$ and $0 < U < \infty$.
$W_{\rm eff}=\sqrt{2t_1^2+2t_2^2}$ is the effective band width of the free DOS.
The orange line (``Nagaoka'') marks the critical interaction strength below which the fully polarized state becomes instable against the paramagnet.
The green line (``HF/Stoner'') is the phase boundary according to the Stoner criterion.
}
\labfig{liebmattis}
\end{figure}

Ground-state ferromagnetism in the one-dimensional Hubbard model is restricted by the Lieb-Mattis theorem \cite{LM62} which excludes a finite order parameter in case of  nearest-neighbor hopping only, irrespective of the interaction $U$. 
Including a next-nearest-neighbor hopping $t_2$, however, ferromagnetism is proven to exist for $U=\infty$ in the limit $t_2 \to 0$ ($t_2<0$) for all densities.\cite{UKN94,STUR92,LCH94}
This limit has to be contrasted to the limit $t_1 = 0$, but finite $t_2$ (two-chain model) where the Lieb-Mattis theorem applies again.
In the low density limit, the ground state is ferromagnetic for $t_2<-1/4$ at $U=\infty$. \cite{MH95}
With a finite next-nearest-neighbor hopping, ferromagnetism occurs in a rather large
part of the $U$-$n$ phase diagram as has been demonstrated by DMRG calculations of
Daul and Noack. \cite{DN97,DN98} 

To study the effect of local and of short-range non-local quantum fluctuations on the stability of the ferromagnetic ground state and to test the predictive power of mean-field and cluster mean-field approaches, we have applied the dynamical impurity approximation (DIA) and the variational cluster approximation (VCA) (see \reffig{refsys}) to the model with $t_2 \ne 0$. 
Note that a finite $t_2$ translates into an asymmetric free DOS and that $t_2 \ne 0$ implies magnetic frustration with respect to antiferromagnetic order.
We exclusively consider $t_2<0$ ($t_1=1$ sets the energy scale) which implies that ferromagnetic order is expected to show up for fillings below half-filling. 

We first check whether or not the Lieb-Mattis theorem is respected by the most simple DIA.
To this end, DIA calculations have been performed to map out the $U$-$t_2$ phase diagram at a fixed filling $n=0.5$ (quarter filling).
Calculations are done using chains with up to 8000 sites. 
The critical interaction for ferromagnetic order $U_c$ is determined by the divergence of the paramagnetic susceptibility.
$\chi$ is calculated by using a finite but small external field ($B=0.01$). 
The values for $U_c$ obtained in this way are checked for selected $t_2$ by calculating the external field for a given small magnetization ($m=0.01$).
Deviations are small, i.e.\ invisible on the scale of the figures discussed below, and can be neglected.

The resulting phase diagram is shown in \reffig{liebmattis}.
The DIA predicts a $U_c$ which varies strongly with $|t_2|$.
To be able to display the results for $0<U<\infty$ and $-\infty< t_2 < 0$ in a single picture, non-linear scales for $t_2$ and $U$ have been used.
The effective bandwidth defined as $W_{\rm eff} = \sqrt{2t_1^2+2t_2^2}$, i.e.\ the standard deviation of the free DOS, and $t_1$ are chosen as the relevant scales for $U$ and $t_2$.

\begin{figure}
\centerline{\includegraphics[width=.85\columnwidth,clip=]{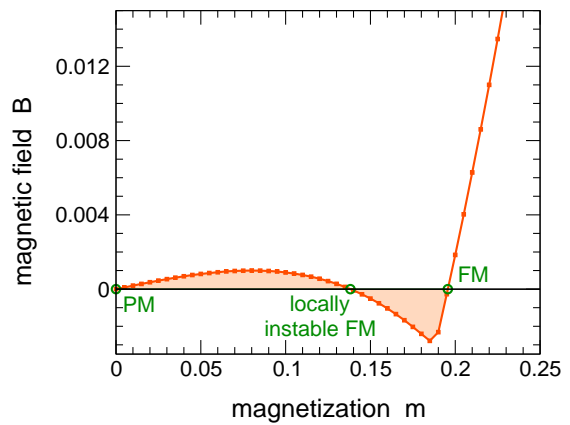}}
\caption{
Maxwell construction for the discontinuous phase transition for $n=0.5$ and $t_2 = -0.34$ ($|t_2|/(|t_1|+|t_2|) \approx 0.25$) at the critical interaction $U_c=0.43$. 
The shaded areas have the same size, i.e.\ $\Delta G = 0$ (see \refeq{DeltaG}).
There is a solution with spontaneous ferromagnetic order at $m=0.20$.
The second ferromagnetic solution at $m\approx 0.14$ shows a negative susceptibility $\chi = \partial m / \partial B < 0$ and is thus locally (and globally) unstable.
}
\labfig{firstorder1}
\end{figure}

\begin{figure}
\centerline{\includegraphics[width=.99\columnwidth,clip=]{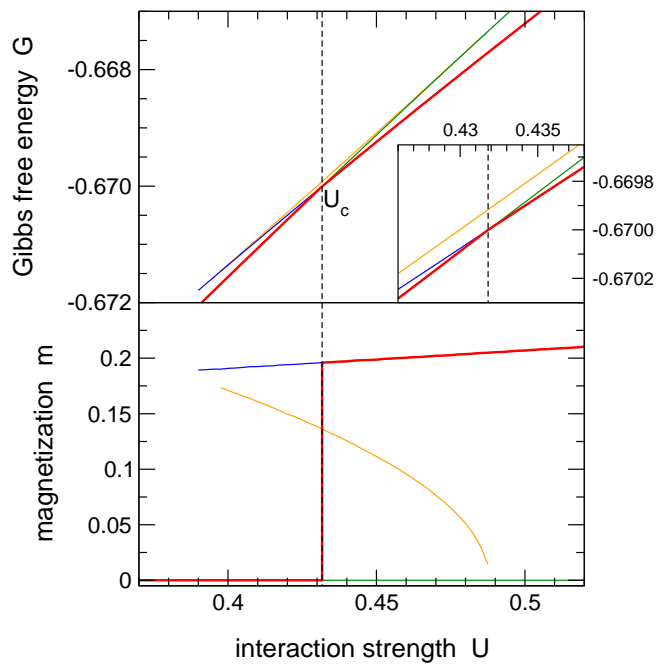}}
\caption{
Gibbs free energies $G$ of the paramagnetic and of the ferromagnetic solutions (top panel) and the corresponding magnetizations $m$ (bottom) as functions of the interaction strength $U$ at $t_2=-0.34$ and quarter filling. 
The red line shows the actual course of the stable solution. 
The magnetization vanishes continuously at $U=0.49$, and at the same interaction strength the susceptibility $\chi$ diverges.
The true phase transition is discontinuous and takes place at $U_c=0.43$.
}
\labfig{firstorder2}
\end{figure}

In both cases where the Lieb-Mattis theorem holds, for $t_2 = 0$ and $t_2 \to \infty$ the DIA predicts $U_c \to \infty$, i.e.\ the absence of ferromagnetic order.
On the other hand, the static mean-field theory is clearly at variance with the exact theorem as can be seen from \reffig{liebmattis} where the green line (``HF/Stoner'') displays the divergence of the Hartree-Fock susceptibility as determined from the Stoner criterion. 
Note that the discrepancy between the DIA and the HF results is drastic except in the vicinity of $t_2 \approx -0.3$ ($|t_2|/(|t_1|+|t_2|) \approx 0.23$) where the chemical potential of the non-interacting system coincides with a van Hove singularity in the free DOS.
Here the Stoner criterion correctly predicts the ferromagnetic instability of the ground state.

As one cannot expect that the Lieb-Mattis theorem is respected rigorously within a mean-field approach, we also performed calculations for different fillings.
In fact, for $n=0.4$ a divergence of the susceptibility is found for $t_2=0$. 
However, the large critical value for the interaction, $U_c \approx 30$, indicates that the violation of the Lieb-Mattis theorem is ``weak'' in the sense that it occurs for extremely strong interactions only. 

The comparison with Hartree-Fock theory tells us that local quantum fluctuations are very important.
On the other hand, the comparison with VCA results shows that non-local fluctuations are not important in first place.
For example, for $t_2=0$ and for a very strong interaction, $U=10^4$, we find a divergence of the susceptibility at a critical filling $n=0.49$ within the DIA while $n=0.42$ within the VCA for $L_c=2$. 
This is the correct trend as the critical filling must vanish for $L_c\to \infty$ due to the Lieb-Mattis theorem.
As compared to the improvement of the DIA with respect to static mean-field theory, however, this appears as marginal. 

In most cases we find the phase transition to be discontinuous. 
\reffig{firstorder1} gives an example. 
Here the homogeneous magnetic field $B$ is shown as a function of the magnetization $m$ for
$t_2=-0.34$ and $U=0.43$.
This corresponds to $|t_2|/(|t_1|+|t_2|) \approx 0.25$ and $U/(U+W_{\rm eff}) \approx 0.22$. Spontaneous ferromagnetism requires a finite order parameter $m$ at $B=0$. 
There are three solutions: (i) the paramagnetic state at $m=0$ which shows a positive susceptibility $\chi = \partial m / \partial B$, 
(ii) a thermodynamically unstable ferromagnetic solution with negative $\chi$, and (iii) a stable ferromagnetic solution with $\chi >0$ and $m=0.20$.
The point $|t_2|/(|t_1|+|t_2|) \approx 0.25$ and $U/(U+W_{\rm eff}) \approx 0.22$ is just the transition point as can be seen from the area under the $B(m)$ curve, i.e.\ from the Maxwell construction, but is somewhat {\em below} the point which is plotted in \reffig{liebmattis} and at which the susceptibility $\chi$ diverges: $U_c/(U_c+W_{\rm eff}) \approx 0.24$.

\begin{figure}
\centerline{\includegraphics[width=.95\columnwidth,clip=]{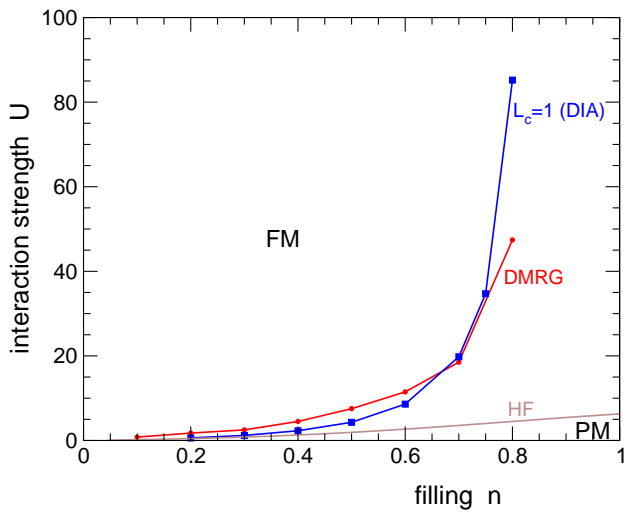}}
\caption{
Magnetic ground-state phase diagram of the one-dimensional Hubbard model for $t_2=-0.2$ as obtained by the DIA. 
DMRG data from Daul and Noack \cite{DN98} and Hartree-Fock (HF) results are shown for comparison.
}
\labfig{cluster}
\end{figure}

This explains itself in \reffig{firstorder2} where the Gibbs free energies and the magnetizations of the three solutions are shown for fixed $t_2$ as a function of $U$.
Note that for $T=0$ and $B=0$ the Gibbs free energy is the ground-state energy: $G=E$.
It can be seen that the thermodynamically unstable solution always has the highest Gibbs free energy.
The actual phase transition therefore takes place at the interaction strength $U=0.43$ where the Gibbs free energies of the (stable) ferromagnetic and of the paramagnetic solutions are crossing.
This is consistent with the Maxwell construction in \reffig{firstorder1} since the 
difference in the Gibbs free energies of the paramagnet and the ferromagnet is
given by 
\be
\Delta G = G_{\rm FM} - G_{\rm PM} = \int_{\rm PM}^{\rm FM} B(m)\, {\rm d}m \; .
\labeq{DeltaG}
\ee
On the phase boundary $\Delta G = 0$.
The divergence of $\chi$, however, is related to the continuous vanishing of the order parameter of the unstable solution at $U_c=0.49$ (see \reffig{firstorder2}, bottom).
Note, that on the scale used in \reffig{liebmattis} the difference between $U_c$ and the true (first-order) transition points is almost invisible.

The $t_2$-range in which the transition is continuous is marked as the shaded area in \reffig{liebmattis}. 
In addition, the figure shows the $t_2$ dependence of the interaction strength at which the fully polarized (``Nagaoka'') state becomes unstable as compared to the paramagnetic state. 
This line crosses the phase transition line (diverging $\chi$) at $|t_2| \approx 0.10$ (corresponding to $|t_2|/(|t_1|+|t_2|) \approx 0.09$) and $|t_2| \approx 0.39$ (corresponding to $|t_2|/(|t_1|+|t_2|) \approx 0.28$).
This implies that for $|t_2|>0.39$ and for $|t_2|<0.09$ the first-order transition is a transition from the paramagnetic to the fully polarized ferromagnetic state while in all other cases the magnetization jumps to a non-saturated value at the respective transition point.
Our DIA results are consistent with the DMRG calculations of Daul \cite{Dau00} which yield a second-order transition at $t_2 = -0.2$ and a first order transition at $t_2 = -0.8$ for quarter filling.

\begin{figure}
\centerline{\includegraphics[width=.95\columnwidth,clip=]{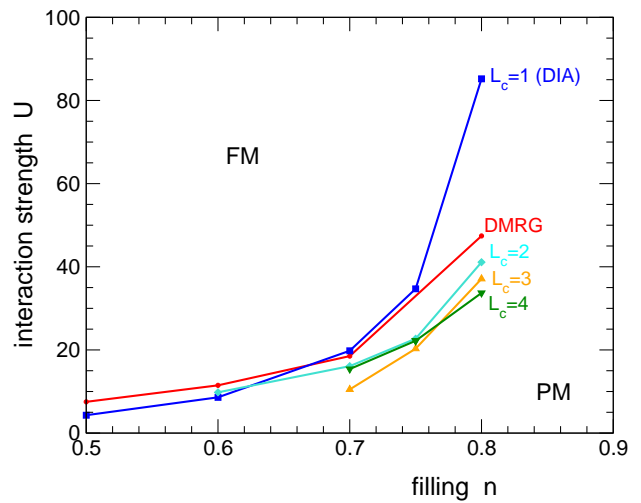}}
\caption{
The same as \reffig{cluster} but using the variational cluster approach with reference systems shown in \reffig{refsys} with $1\le L_c \le 4$ (and the same number of bath sites, i.e.\ $n_{\rm s}=2$) in the range $0.7 < n < 0.8$.
}
\labfig{cluster2}
\end{figure}

For a systematic comparison of the results of the dynamical impurity approach with DMRG data, \cite{DN98} we fix the next-nearest-neighbor hopping to $t_2=-0.2$ and map out the phase diagram $U$ vs.\ filling $n$. 
The result is shown in \reffig{cluster} in comparison with static mean-field theory. 
The phase diagram turns out to be qualitatively similar to the result for infinite dimensions (see \reffig{qmcpd}).
The critical interaction $U_c$ strongly varies with $n$ and becomes extremely large for fillings close to half-filling. 
Despite the simplicity of the reference system, the agreement with the DMRG date is reasonable.

Within static mean-field theory, local magnetic moments are formed in the ferromagnetic state only.
This might be the right picture for the low-density regime. 
For fillings $n \gtrsim 0.5$, however, static mean-field theory fails to reproduce the phase diagram as the tendency towards ferromagnetic order is overestimated drastically.

Opposed to static mean-field theory, the DIA allows for local-moment formation already in the paramagnetic state and captures the correlated mean-field physics of the paramagnetic Mott transition at half-filling. \cite{Pot03b,EKPV07,BKS+09} 
The high values for the Hubbard interaction necessary to produce ferromagnetic order can be understood in a picture where the ferromagnetic state evolves from a highly correlated paramagnet with preformed but disordered local magnetic moments. 
With increasing $U$ and with increasing fillings $n$ the local magnetic moments are more and more efficiently screened by a collective (single-band) Kondo effect.
The tendency to screen the local moments counteracts the formation of magnetic order and thus results in very strong critical interactions.
Such a mechanism is already included in the DIA.
From the reasonable agreement with the DMRG data and the strong improvement with respect to static mean-field theory, we therefore infer that this mechanism is essential.
While the correct Kondo scale cannot be captured with a single bath site ($n_{\rm s}=2$), the possibility to form a local singlet within a thermodynamically consistent approximation appears to be a key ingredient to understand the phase diagram.

Besides a screening of the local moment by local fluctuations, a screening by {\em non-local} fluctuations is conceivable. 
This would lead to non-local singlets -- or even to long-range {\em anti-}ferromagnetic order. 
Consequently, such a mechanism is expected to be effective for fillings close to half-filling where non-local antiferromagnetic correlations are important.
Note, however, that due to the Lieb-Mattis theorem the necessity to include a finite $t_2$ already suppresses antiferromagnetic order by magnetic frustration to some degree. 
This might explain that the effect of non-local fluctuations appears to be comparatively weak for intermediate fillings and significant for fillings close to half-filling only.

This can be seen in \reffig{cluster2} where we compare the DIA phase diagram with the results obtained from VCA calculations with finite clusters as reference systems: $L_c=2-4$
while the description of the local degrees is unchanged ($n_{\rm s}=2$, one bath site per correlated site). 
For $n \lesssim 0.75$ the critical interaction does not change much while for $n=0.8$, the critical $U$ is strongly reduced in the cluster approach.
The VCA thereby improves the agreement with the DMRG.

Although the step from $L_c=1$ to $L_c>1$ appears to be essential close to half-filling, the results of the cluster approach have to be interpreted with some care since the expected convergence with increasing cluster size can hardly be seen for $L_c\le 4$. 
This reflects finite-size errors the size of which can be estimated by comparing the results for different $L_c$ among each other. 
Within this (considerable) error there is agreement with the DMRG results. 
On the other hand the VCA, and more important even the DIA, is able to predict the qualitatively correct trend for the phase diagram. 
It is also important to note within the DIA it is much easier to find, stabilize and trace magnetic solutions. 
For the clusters with $L_c>1$ we have not been able to find solutions in the entire filling range for reasons discussed in Sec.\ \ref{sec:lc}.

\section{Conclusions}
\label{CON}

The self-energy-functional theory has been applied to the Hubbard model in infinite and in one dimension to investigate spontaneous ferromagnetic order. 
Using different reference systems generating different single-site and cluster mean-field approximations, i.e.\ dynamical impurity and the variational cluster approximations, it is possible to study the effects of local and of short-range non-local quantum fluctuations on the stability of the ferromagnetic ground state. 

We find that local fluctuations are of crucial importance to get a qualitatively correct phase diagram and to respect the Lieb-Mattis theorem. 
Opposed to static mean-field theory, ferromagnetic order quite generally requires substantially higher interaction strengths and can be understood as evolving from preformed but disordered local magnetic moments. 
The extremely large critical interactions found in one-dimensional chains could then be attributed to the screening of the local moments by local (Kondo-type) correlations which becomes more and more effective for increasing filling or interaction strength.
This local singlet formation, on a qualitative level, is already included in the DIA.
Singlet formation due to non-local correlations, included in the VCA, appears to be relevant for fillings close to half-filling only, while a ferromagnetic ground state can be obtained in large areas of the parameter space and down to the low-density limit in particular. 
The limited importance of (antiferromagnetic) non-local correlations is of course interrelated with the frustration of antiferromagnetic order due to a next-nearest-neighbor hopping $t_2$, with the Lieb-Mattis theorem, and, in the case of infinite dimensions, with an asymmetric free DOS.
There is an obvious similarity of the magnetic phase diagram of the one-dimensional model with the phase diagram in infinite dimensions which again suggests that local correlations play the predominant role for ferromagnetic order.

This is of some importance for future investigations of ferromagnetism in nano-sized objects, e.g.\ chains or clusters on non-magnetic substrates, as it opens a route to study those systems by (dynamical) mean-field methods which, as concerns the system geometry, are more flexible than the density-matrix renormalization group, for example.
The situation may be contrasted, e.g., with the absence of long-range antiferromagnetic order in one-dimensional systems which is caused by non-local quantum fluctuations. 
In the latter case any mean-field approach would be questionable a priori.  

The study of infinite-dimensional lattices using the DIA has shown that the previously known parameter ranges that are favorable for ferromagnetic order, namely low to intermediate fillings and moderate $U$ in case of a strongly asymmetric free DOS and fillings close to half-filling and extremely strong $U$ in case of a symmetric free DOS, are linked continuously. This demonstrates the difficulty to find simple ``mechanisms'' for ferromagnetic order in the Hubbard model.
Ferromagnetism should therefore be seen as a complex phenomenon the description of which necessarily requires non-perturbative and thermodynamically consistent many-body techniques.

To study ferromagnetism within self-energy-functional theory, at least six variational parameters have to be optimized simultaneously, namely the spin-dependent one-particle energies of the correlated and of the uncorrelated sites and the spin-dependent hybridization strength in addition. 
Two more variational parameters must be considered for calculations at fixed filling and magnetization which is convenient for the construction of phase diagrams.
This can be accomplished by a number of technical improvements concerning (i) an accurate treatment of k- and frequency summations, (ii) optimization algorithms which adapt to the local structure of the functional, (iii) global optimization algorithms to find a stationary point of the functional.
For the calculation of the static and homogeneous paramagnetic susceptibility, an optimization of spin-independent parameters is sufficient.

The comparison with dynamical mean-field theory and with density-matrix renormalization-group calculations for the infinite-dimensional and for the one-dimensional model, respectively, has been essential to rate the approximations.
For both, infinite dimensions and one dimension, a simple DIA turns out to be sufficient for a qualitative and rough scan of the phase diagram.
This might be sufficient in view of the fact that the Hubbard and similar models themselves represent strong simplifications as compared to a real material.
In one dimension, a cluster approach including short-range correlations, i.e.\ the VCA, appears to be necessary for fillings close to half-filling.
A satisfactory convergence with increasing cluster size, however, could not be obtained. 
For future studies of more complicated low-dimensional geometries, we therefore suggest to use the DIA in those ranges of the parameter space where there are no significant deviations from results obtained by the cluster approach.

\acknowledgments
We would like to thank Robert Eder (Karlsruhe Institute of Technology) for discussions. 
Support of this work by the Deutsche Forschungsgemeinschaft within the Sonderforschungsbereich 668 (project A14) and by the Landesexzellenzinitiative Hamburg ``Nanospintronics'' is gratefully acknowledged.


\end{document}